\documentclass[aps,prd,groupedaddress,nofootinbib,twocolumn]{revtex4}
\usepackage{graphicx,amssymb,amsmath,bm}
\usepackage{dsfont} 
\usepackage[caption=false]{subfig}
\usepackage{multirow}

\newcommand{\be}{\begin{equation}}
\newcommand{\ee}{\end{equation}}
\newcommand{\bea}{\begin{eqnarray}}
\newcommand{\eea}{\end{eqnarray}}
\newcommand{\nab}{\nabla}
\newcommand{\nn}{\nonumber}
\def\rf#1{(\ref{#1})}

\usepackage{accents}
\newlength{\dhatheight}

\newcommand{\bra}[1]{\left(#1\right)}

\newcommand{\brac}[1]{\left\{#1\right\}} 
\newcommand \veps {\varepsilon}
\newcommand{\A}{{\cal A}}
\newcommand{\E}{{\cal E}}
\renewcommand{\H}{{\cal H}}
\newcommand{\K}{{\cal K}}
\newcommand{\M}{\mathbb{M}}
\newcommand{\PP}{\mathbb{P}}
\begin{document}

\title{The covariant Tolman-Oppenheimer-Volkoff equations II: The anisotropic case}

\author{Sante Carloni}
\author{Daniele Vernieri}
\address{ Centro Multidisciplinar de Astrof\'{\i}sica - CENTRA,
Instituto Superior Tecnico - IST,
Universidade de Lisboa - UL,
Avenida Rovisco Pais 1, 1049-001, Portugal}
\begin{abstract}
We generalise the covariant Tolman-Oppenheimer-Volkoff equations proposed in Ref.~\cite{SDTOV1} to the case of static and spherically symmetric spacetimes with anisotropic sources. The extended equations allow a detailed analysis of the role of the anisotropic terms in the interior solution of relativistic stars and lead to the generalisation of some well known solutions of this type. We show that, like in the isotropic case, one can define generating theorems for the anisotropic Tolman-Oppenheimer-Volkoff  equations. We also find that it is possible to define a reconstruction algorithm able to generate a double infinity of interior solutions. Among these, we derive a class of solutions that can represent ``quasi-isotropic'' stars.
\end{abstract}
\maketitle

\tolerance=5000
\section{Introduction} 
It has been clear for a long time that, differently from the Newtonian case, in General Relativity the structure of a spherical stellar object can be highly non trivial. Indeed, in spite of the progress made so far, no completely satisfactory general solution of this problem has been found.

The isotropic case, in which the source of the gravitational field is a perfect fluid, was the focus of much research in the last few years. For example, in Ref.~\cite{Delgaty:1998uy} an extensive selection of the known solutions was made using physical criteria like the positivity of the pressure and the negative gradient of the density distribution. In addition, a number of new methods to obtain analytic solutions representing relativistic stars  were proposed (e.g. Ref.~\cite{Mak:2005eq,Lake:2002bq,Mak:2013pga}), together with an analysis  of the general properties of these solutions~\cite{Baumgarte:1993}. In Refs.~\cite{Boonserm:2005ni,Boonserm:2006up} {\it e.g.} some general theorems were proven that connect different isotropic solutions to each other.

On the other hand, actual relativistic astrophysical stellar objects hardly resemble spheres of perfect fluids. This is due to a number of reasons. For example magnetic fields can be very intense and induce non trivial deviations from isotropy. In addition, it is widely believed that in relativistic stellar objects matter is in a state which has exotic thermodynamical properties (see e.g. Refs.~\cite{Herrera:1997plx,Harko:2002db} and references therein). In fact, it has been suggested that even a simple mixture of real gas can behave as an anisotropic fluid Ref.~\cite{Letelier:1980mxb}.

The study of solutions describing anisotropic stellar interiors presents a number of additional challenges, due to the extra degrees of freedom associated with the anisotropy. There are a number of interesting works in which solutions representing anisotropic stars are proposed and analysed from different points of view (see {\it e.g.} Refs.~\cite{Gupta2000abc,Dev:2000gt,Hernandez:2001nr} for some recent examples). 

In Ref.~\cite{SDTOV1}, we developed a new covariant formalism to treat the Tolman-Oppenheimer-Volkoff (TOV) equations in the case of an isotropic fluid. In the new formalism many aspects of the properties of these solutions, like the mathematical structure of the equations, become immediately clear. In addition, the generating theorems mentioned above, plus some new ones, can be easily formulated as linear deformations of the initial solutions. The formalism also allows the determination of a number of physically relevant solutions via the use, for example,  of reconstruction algorithms.

The purpose of this paper is to extend this formalism to the case of anisotropic sources. As in the isotropic case, the new equations will clarify the mathematical structure of the problem and suggest in an intuitive way a number of purely analytical resolution strategies. The characterisation of the  anisotropy as a pressure term, will also allow the definition of a new class of generating theorems for this case. We will also prove, via the definition of a reconstruction algorithm, that, surprisingly, generating solutions in the anisotropic case is indeed easier than the isotropic one.

The paper is organised as follows. Section II introduces the basic formalism and a key set of variables which will be useful for our purposes. Section III is dedicated to the construction of the covariant TOV equations and to different resolution strategies of the new equations. In Section IV we formulate some generating theorems of the anisotropic case. In Section V instead we will propose a reconstruction algorithm and we will use it to generate some physically interesting solutions. Section VI is dedicated to the Conclusions.

Unless otherwise specified, natural units ($\hbar=c=k_{B}=8\pi G=1$) will be used throughout this paper and Latin indices run from 0 to 3. The symbol $\nabla$ represents the usual covariant derivative and $\partial$ corresponds to partial differentiation. We use the $-,+,+,+$ signature and the Riemann tensor is defined by
\begin{equation}
R^{a}{}_{bcd}=\Gamma^a{}_{bd,c}-\Gamma^a{}_{bc,d}+ \Gamma^e{}_{bd}\Gamma^a{}_{ce}-\Gamma^e{}_{bc}\Gamma^a{}_{de}\;,
\end{equation}
where the $\Gamma^a{}_{bd}$ are the Christoffel symbols ({\it i.e.} symmetric in the lower indices), defined by
\begin{equation}
\Gamma^a_{bd}=\frac{1}{2}g^{ae}
\left(g_{be,d}+g_{ed,b}-g_{bd,e}\right)\;.
\end{equation}
The Ricci tensor is obtained by contracting the {\em first} and the {\em third} indices
\begin{equation}\label{Ricci}
R_{ab}=g^{cd}R_{acbd}\;.
\end{equation}
Symmetrisation and the anti-symmetrisation over the indices of a tensor are defined as 
\begin{equation}
T_{(a b)}= \frac{1}{2}\left(T_{a b}+T_{b a}\right)\;,~~ T_{[a b]}= \frac{1}{2}\left(T_{a b}-T_{b a}\right)\,.
\end{equation}
Finally the Hilbert--Einstein action in the presence of matter is given by
\begin{equation}
{\cal A}=\frac12\int d^4x \sqrt{-g}\left[R+ 2{\cal L}_m \right]\;.
\end{equation}

\section{$1+1+2$ Covariant approach in brief} 

A more general and detailed presentation of this formalism can be found in Refs.~\cite{Clarkson:2002jz,Betschart:2004uu,Clarkson:2007yp} and in Ref.~\cite{SDTOV1}. We give here a very brief description of its main features.

The $1+1+2$ approach is based on a double foliation of the spacetime: a given manifold is first foliated in spacelike 3-surfaces and then these 3-surfaces are foliated in 2 surfaces. These foliations are obtained by defining a congruence of integral curves of the time-like $(u_{a} u^{a} = -1)$ vector field $u^a$ and a congruence of  integral curves defined by the spacelike $(e_{a} e^{a} = 1)$ vector $e_a$. 

The $u^a$ and  $e_a$ congruences allow the definition of two projection tensors:  $h^a{}_b=g^a{}_b+u^a u_b$ which represents the metric of the 3-spaces ($S$)  orthogonal to $u_a$ and  
\be 
N_{a}{}^{b} \equiv h_{a}{}^{b} - e_{a}e^{b} = g_{a}{}^{b} + u_{a}u^{b} 
- e_{a}e^{b}~,~~N^{a}{}_{a} = 2~, 
\label{projT} 
\ee 
which represents the metric of the 2-spaces ($W$) orthogonal to $u_a$ and $e_a$.

The 2-surface $W$ carries a natural 2-volume element, the Levi-Civita 2-tensor: 
\be
\veps_{ab}\equiv\veps_{abc}e^{c} = \eta_{dabc}e^{c}u^{d}~, \label{perm}
\ee 
where $\veps_{abc}$  and $\eta_{abcd}$ are the  Levi-Civita tensors in $S$ and the original manifold respectively. 

The decomposition defined above can be also used to define a set of derivatives operators:
\bea
&&\dot{X}^{a..b}{}_{c..d}{} \equiv  u^{e} \nab_{e} {X}^{a..b}{}_{c..d}~, \\
&& D_{e}X^{a..b}{}_{c..d}{} \equiv  h^a{}_f h^p{}_c...h^b{}_g h^q{}_d h^r{}_e \nab_{r} {X}^{f..g}{}_{p..q}~,\\
&& \hat{X}_{a..b}{}^{c..d} \equiv  e^{f}D_{f}X_{a..b}{}^{c..d}~,  \\
&&\delta_fX_{a..b}{}^{c..d} \equiv  N_{a}{}^{f}...N_{b}{}^gN_{h}{}^{c}..
N_{i}{}^{d}N_f{}^jD_jX_{f..g}{}^{i..j}\;, 
\eea
which  are called \textit{covariant time derivative},  \textit{orthogonally projected covariant derivative}, \textit{hat-derivative}, \textit{$\delta$ -derivative}, respectively.

The kinematic variables of the $1+1+2$ formalism are eight scalars
\begin{equation}
\brac{\phi, \xi, \Theta, \A, \Omega,\Sigma, \E, \H},
\end{equation}
six vectors
\begin{equation}
\brac{a^b,\A^{a},\Omega^{a}, \Sigma^{a}, \E^{a}, \H^{a}},
\end{equation}
and four tensors
\begin{equation}
\brac{\zeta^{ab},\Sigma^{ab}, \E^{ab}, \H^{ab}}.
\end{equation}
These quantities are defined via the decomposition of  the covariant derivative of $u_a$, the orthogonally projected  derivative of  $e^a$ and of the Weyl tensor $C_{abcd}$ into their irreducible parts. More specifically:
\be
\begin{split}
& \A=e_a \dot{u}^{a}~, \quad  \A^a=N_{ab}\dot{e}^{b}~,\\
& \tilde{\theta}=\delta_a u^a~,\quad \bar{\theta}=a_{b}u^b,\quad\tilde{\theta}+\bar{\theta}=D_a u^a=\Theta~,\\
&\Omega =\frac{1}{2}\veps^{abc}D_{[a} u_{b]} e_a~, \quad\Omega^{a}=\frac{1}{2}\veps^{abd}D_{[a} u_{b]} N_d{}^{a}~,\\
&\Sigma=\sigma^{ab}\bra{ e_ae_b - \frac{1}{2}N_{ab}}~,\quad \Sigma_a=\sigma_{cd} e^c N^{d}{}_{a}~,\\
&\Sigma_{ab}= \bra{ N^{c}{}_{(a}N_{b)}{}^{d} - \frac{1}{2}N_{ab} N^{cd}} \mathds{\sigma}_{cd}~,\\
&\sigma_{ab}= \left(n^{c}{}_{(a}h_{b)}{}^{d} - \frac{1}{3}h_{ab} h^{cd}\right)D_{c}u_{d}~,\\
&a_{b} =  e^{c}{\rm D}_{c}e_{b} = \hat{e}_{b}~, \qquad \phi = \delta_a e^a~, \\  
&\zeta_{ab} =  \bra{ N^{c}{}_{(a}N_{b)}{}^{d} - \frac{1}{2}N_{ab} N^{cd}} \delta_{ c}e_{d}~, \\
&\xi =  \frac{1}{2} \veps^{ab}\delta_{a}e_{b}~,
\end{split}
\ee
and 
\be
\begin{split}
&\E=C^{ab}{}_{cd}u^{c}u^{d}\bra{ e_ae_b - \frac{1}{2}N_{ab}}~,\\
&\E_a=C_{cd ef}u^{e}u^{f} e^c N^{d}{}_{a}~,\quad \E_{ab}= C_{\{ab\}cd}u^{c}u^{d},\\
&\H=\frac{1}{2}\veps^{a}{}_{de}C^{de b}{}_{c}u^c\bra{ e_ae_b - \frac{1}{2}N_{ab}}~,\\
&\H_a=\frac{1}{2}\veps_{cfe}C^{fe}{}_{dh}u^h e^c N^{d}{}_{a}~,\\
&\H_{ab}=\frac{1}{2}\veps_{\{a}{}^{de}C_{b\}cde}u^c~.
\end{split}
\ee

For an observer that moves on the geodesic congruence defined by $u_a$, the expansion  of the geodesics will be  given by $\Theta=\tilde{\theta}+\bar{\theta}~$, the deviation from free fall will be represented by the  components $\A$  and $\A^{a}$ of  the acceleration vector $\dot{u}_a$,  the components $\Sigma_{ab}$, $\Sigma_{a}$ and $\Sigma$ of the  shear $\sigma_{ab}$ will represent the non isotropic deformation of the geodesic flow and the components  $\Omega$ and $\Omega_a$ of the vorticity $\omega_{ab}=D_{[a} u_{b]}$ its rotation. Similarly if the same observer chooses $ e^{a} $ as special direction in the spacetime, $\phi$ represents the expansion of the integral curves of the vector field $ e^{a} $,  $\zeta_{ab}$ is their distortion  ({\it i.e.} the \textit{shear of $e^{a}$}) and $a^{a}$ the change of the vector $e_a$ along its integral curves ({\it e.g.} its \textit{acceleration}). We can also interpret $\xi$ as a representation of the ``twisting'' or rotation of the integral curves of $e_a$ ({\it i.e.} the \textit{vorticity} associated with $e^{a}$). Finally $\E, \E_a, \E_{ab}$ are related with the Newtonian part of the gravitational potential while $\H, \H_a, \H_{ab}$ are related to tidal relativistic forces.

The matter sources are represented by four scalars: 
\begin{equation}
\brac{\mu, p, Q,\Pi}~,
\end{equation}
which denote the standard matter density and pressure  and the scalar part of the heat flux and the anisotropic pressure; two vectors
\begin{equation}
\brac{Q^{a},\Pi^{a}},
\end{equation}
which denote  the solenoidal part of the heat flux and the anisotropic pressure; and the tensor $\Pi^{ab}$. The definitions are
\be
\begin{split}
& \mu=T_{ab}{u}^{a}{u}^{b}~,\\
&p=\frac{1}{3}T_{ab}\left({e}^{a}{e}^{b}+N^{ab}\right)~,\\
& \Pi=\frac{1}{3}T_{ab}\left(2{e}^{a}{e}^{b}-N^{ab}\right)~,\\
&Q=\frac{1}{2}T_{ab}e^{a}u^{b}~,\\
&Q_a=T_{cd}N^{a}{}_{c}u^{d}~,\\
&\Pi_{a}=T_{cd}N^{a}{}_{c}e^{d}~,\\
&\Pi_{ab}=T_{\{ab\}}~.
\end{split}
\ee
The three matter sources $\brac{\mu, p_r,p_\perp}$ that appear in the temporal, radial and angular parts of the Einstein equations 
can be directly connected to  the $1+1+2$ matter scalars. In particular, 
 \begin{align}
& p_r= p+\Pi,\\
& p_\perp=p-\frac{1}{2}\Pi.
\end{align}

The formalism given above is able to describe in a natural way all locally rotationally symmetric (LRS) {\it i.e.} spacetimes in which one can define covariantly a unique, preferred spatial direction. In the following we are interested in the case of the rotation free, static and spherically symmetric spacetimes (LRSII). In this case \textit{all} the $1+1+2$ vectors and tensors vanish as well as the variables $\Omega$, $ \xi $, $ \H $, $\Theta$, $\Sigma$ and $Q$. Thus one is left with the six scalars $\brac{\A, \phi,  \E,  \mu, p, \Pi }$
which are related by the equations
\bea 
\hat\phi &= -&\frac12\phi^2 -\frac23\mu-\frac12\Pi-\E~,
\label{StSpSymEqGen1}\\
\hat\E -\frac13\hat\mu + \frac12\hat\Pi &=-& \frac32\phi\bra{\E+\frac12\Pi}~,
\label{StSpSymEqGen2}\\
\hat p+\hat\Pi&= -&\bra{\frac32\phi+\A}\Pi-\bra{\mu+p}\A~,
\label{StSpSymEqGen3}\\
\hat\A &= -&\bra{\A+\phi}\A + \frac12\bra{\mu +3p}~, \\
\label{StSpSymEqGen4}
\hat{K} &=& -\phi K~,
\label{StSpSymEqGen5}
\eea
with the constraints
\be
\begin{split}
&0 = - \A\phi + \frac13 \bra{\mu+3p} -\E +\frac12\Pi,\\
&K = \frac13 \mu - \E - \frac12 \Pi + \frac14 \phi^{2}~.
\end{split}
\ee

We introduce at this point an affine parameter such that the Gaussian curvature $K$ is given by~\cite{Carloni:2014rba} 
\begin{equation}
K= K_0 e^{-\rho}
\end{equation}
The parameter $\rho$ is connected to the standard area radius by  
\begin{equation}\label{RhoR}
\rho=2\ln(r/r_0),
\end{equation}
where $r_0$ is an arbitrary constant. In the rest of this work we will perform the calculations in $\rho$, but we will give the final results in terms of $r$ to make the comparison with known results easier.  

Using  the parameter $\rho$ and defining the following variables~\cite{Carloni:2014rba}
\begin{eqnarray}
&&\nn X=\frac{\phi_{,\rho}}{\phi}\,,\qquad Y= \frac{\A}{\phi}\,, \qquad \K=\frac{K}{\phi^2}\,, \qquad E=\frac{\E}{\phi^2}\,, \\
&& \mathbb{M}=\frac{\mu}{\phi^2}\,, \qquad \mathds{P}=\frac{\Pi}{\phi^2}\,, \qquad P=\frac{p}{\phi^2}\,,\label{Var}
\end{eqnarray}
the Eqs.~(\ref{StSpSymEqGen1})-(\ref{StSpSymEqGen4}) take the form
\bea
&& Y_{,\rho}=M+3 P-2Y (X+ Y+1)\,,  \label{NewEqGenV1}\\
&& \K_{,\rho}=-\K(1+2X)\,, \label{NewEqGenV2}\\
&& \nn P_{,\rho}+\mathds{P}_{,\rho}=-2 Y (M+\mathds{P})-2 P(2X+Y)\\
&&~~~~~~~~~~~~~~~~-\mathds{P} (4 X+3)\,,
 \label{NewEqGenV3}
\eea
with the constraints
\begin{eqnarray}
&& 2 \mathbb{M}+2 P+2 \mathds{P}+2X-2 Y+1=0\,,\label{NewConstrGenV1}\\
&&1- 4\K-4P+4Y-4\mathds{P}=0\,, \label{NewConstrGenV2}\\
&&2 \mathbb{M}+6 P+3 \mathds{P}-6 Y-6 E=0\,. \label{NewConstrGenV3}
 \end{eqnarray}
These equations will be the starting point for the construction of the covariant TOV equations.

Finally, it is useful to give the form of $\A$, $\phi$, $K$ , $Y$ and $\K$ in terms of the the metric coefficients in  and their derivatives. In $\rho$, we have
\begin{equation}
\begin{split}
&\A=\frac{1}{2A\sqrt{B} }\frac{d A}{d \rho}\,, \qquad
\phi=\frac{1}{\sqrt{B}}\,,\\ 
&K=\frac{1}{C_0} e^{-\rho}=K_0 e^{-\rho}\,,
\label{APhiKsolutionRho}
\end{split}
\end{equation}
\be
\begin{split}
Y=\frac{1}{2}\frac{A_{,\rho}}{A}\,, \quad
\K=K_0 B(\rho) e^{-\rho}\,.
\label{YKaRho}
\end{split}
\ee
Instead in terms of $r$ the same variables read 
\begin{equation}
\begin{split}
&\A=\frac{1}{2 \sqrt{B} }\frac{d A}{d r}~, \qquad
\phi=\frac{2}{r\sqrt{B}}~,\\ 
&K=\frac{\bar{K}_0}{r^{2}}\,,
\label{APhiKsolutionR}
\end{split}
\end{equation}
\be
\begin{split}
Y=\frac{1}{4}\frac{A_{,r}}{A}\,, \qquad
\K=\bar{K}_0 B(r)\,.
\label{YKar}
\end{split}
\ee
It is clear that $r$ will be the classical area radius only if $\bar{K}_0=1$. We will make this assumption from now on. In addition Eq.~\eqref{RhoR} implies that $K_0= r_0^{-2}$. In the following we will use this relation any time that we will write the solution in terms of $r$.
\subsection{Junction Conditions} 
Sometimes  a non vacuum static spherically symmetric metric has interesting properties, but its asymptotic properties  are not useful to model stellar objects. In this case such solution can be matched to the vacuum Schwarzschild solution at finite values of the radial parameter. There is a set of well known conditions to be satisfied in order to perform such soldering, {\it i.e.} the  Israel's junction conditions~\cite{Israel:1966rt,Barrabes:1991ng}. We will now deduce the equivalent of these conditions in terms of the new variables. The procedure is analogous to the one in Ref.~\cite{SDTOV1} for the isotropic case. 

The surface of separation will be the 3-surfaces $\mathcal{S}$ normal to $e_a$. Indicating the jump along $\mathcal{S}$ as $[X]=X^+ -X^-$, Israel's conditions in the spherically symmetric case read 
\begin{equation}
\begin{split}
&[g^\mathcal{S}_{ab}]=[N_{ab}+u_{a}u_{b}]= 0\,,\\
& [\bar{K}_{cd}]=\left[\frac{1}{2}\phi N_{ab}-u_au_b \A\right]=0\,,
\end{split}
\end{equation}
where $g^\mathcal{S}_{ab}$ is the metric of $\mathcal{S}$ and $\bar{K}_{cd}$ is the extrinsic curvature of $\mathcal{S}$. The conditions above imply the following junction conditions
\begin{equation}
[u_a]=0\,, \qquad [N_{ab}]=0\,,
\end{equation}
and
\begin{equation}\label{Israel2}
[\A]=0, \qquad [\phi]=0\,.
\end{equation}
If the  conditions \eqref{Israel2} are not satisfied a thin shell with stress-energy tensor  
\begin{equation}\label{IsraelShell}
\begin{split}
&T^\mathcal{S}_{ab}=(N_{ab}+u_{a}u_b)[\bar{K}]-[\bar{K}_{cd}] = \\
&~~~~~=u_au_b[\phi+2 \A] +N_{ab} \left[\frac{\phi}{2}+\A\right],
\end{split}
\end{equation}
will be present in the spacetime. 

Going back to the variables in Eq.~\eqref{Var}, since $[\phi]=0$ then $[\K]=0$, and using the constraint in Eq.~\rf{NewConstrGenV2} above one gets
\begin{equation}
0=[\K]= \left[P+\mathds{P}-Y\right]\,.
\end{equation}
Now, since $[\A]=0$, we have $[Y]=0$ and the junction condition turns out to be
\begin{equation}\label{JunCov}
 \left[P+\mathds{P}\right]=0\,.
\end{equation}
As $[\phi]=0$, this is equivalent to say that $[p_r]=0$. In other words, in order to provide a smooth junction with the Schwarzschild metric,  one has to seek a value of the radius in which the radial pressure is zero.  In the following we will impose the junction conditions requiring directly that $p_r$ is zero on the boundary of the star. We will have, however, two cases in which this condition is incompatible with the structure of the solution and we will be required to introduce a shell. 

The above condition does not give any information on the energy density and the tangential pressure, implying that there is no constraint on these quantities. This can be verified breaking covariance. From  Eqs.~(\ref{APhiKsolutionRho}-\ref{YKaRho}) and~(\ref{APhiKsolutionR}-\ref{YKar}) we realise that the junction conditions \eqref{Israel2} require the continuity of $A$, of its first derivative and also  the continuity of $B$. From the Einstein equations it is easy to see that the energy density and the tangential pressure depend on derivatives of the metric coefficients (like the first derivative of $B$) which have no constraint. As a consequence both of these quantities can have a jump at the junction. 

\section{$1+1+2$ TOV equations}\label{ANISOTOV}
Considering Eq.~\rf{NewEqGenV2} and eliminating $X$ and $Y$ from Eq.~\rf{NewEqGenV3}, one obtains
\begin{equation}\label{112TOVGen}
\begin{split}
&P_{,\rho }+\mathbb{P}_{,\rho }=P \left(\mathbb{M}-2
   \mathbb{P}-3 \mathcal{K}+\frac{7}{4}\right)\\
&~~~~~~~~~~~~~+\mathbb{P} \left(\mathbb{M}-3
   \mathcal{K}+\frac{1}{4}\right)\\
&~~~~~~~~~~~~~+\left(\frac{1}{4}-\mathcal{K}\right)
   \mathbb{M}-P^2-\mathbb{P}^2\,,\\
&\mathcal{K}_{,\rho }=-2 \mathcal{K}\left(\mathcal{K}-\frac{1}{4}-\mathbb{M}\right)\,.
\end{split}
\end{equation}
The structure of these equations is similar to the one of the isotropic case treated in Ref.~\cite{SDTOV1} and therefore we have similar problems in determining a general analytical solution. The most important difference is that now two different pressure terms (isotropic and anisotropic) appear in the TOV equations. In the standard treatment, the equations above are written in a slightly different way. Indeed, the TOV equations are written for the combination $P+\mathbb{P}$ which would correspond to $p_r/\phi^2$ (see {\it e.g.} Ref.~\cite{Herrera:1997plx}). We will make use of the latter form of the equations to give a special case in the following subsection. 

In general, since anisotropic solutions posses additional degrees of freedom, it is natural to expect that this case is more complicated that the isotropic one. Surprisingly we will find that this is not always the case.

The equations above are completely equivalent to the Einstein equations. However, not all the solutions of the Einstein equations with anisotropic fluids correspond to (the interior of) stellar objects. It is known~\cite{Herrera:1997plx,Mak:2013pga} that a physical solution will have to fulfil the following conditions
\begin{enumerate}
\item $\mu$,  $p_r$ and $p_\perp$ should be positive inside the object;
\item the gradients of $\mu$,  $p_r$ and $p_\perp$ should be negative;
\item the speed of sound should be always  less that the speed of light $0<\frac{\partial p_r}{\partial \mu}<1$, $0<\frac{\partial p_\perp}{\partial \mu}<1$; 
\item the energy conditions should be satisfied;
\item the anisotropy $\Pi$ should be zero in the centre of the object, {\it i.e.} $p_r=p_\perp$ at the centre.
\end{enumerate}
In the following we will present solutions that are compatible with these conditions at least for one set of their parameters.

\subsection{Some resolution strategies} \label{SolANITOV}
There are several strategies which can be used to obtain solutions of the TOV equations in this case. The simplest one are based on making an ansatz on the behaviour of the anisotropy $\Pi$ and then solving for the corresponding pressure. This is one on the most common approaches in literature  (see e.g.~\cite{Herrera:1997plx}).  

Setting for example
\begin{equation}
 \mathbb{P}=\frac{ \mu _0^2 P_0 (1-h) e^{2 \rho }
   \left(1-\frac{\mu _0 e^{\rho }}{3}\right)^{h/2}}{18\left(1-\frac{\mu _0 e^{\rho }}{3}\right)^2
   \left[\left(1-\frac{\mu _0 e^{\rho }}{3}\right)^{h/2}+3 P_0\right]^2}\,,\
\end{equation}
where $h$ is an arbitrary constant, one obtains
\begin{subequations}
\begin{align}
 &\nn P=\frac{\mu _0 e^{\rho }}{\left[\left(1-\frac{\mu _0 e^{\rho }}{3}\right)^{h/2}+3 P_0\right]^2} \left[ \frac{3 P_0^2 +\left(1-\frac{\mu _0 e^{\rho }}{3}\right)^h}{4(\frac{\mu _0 e^{\rho }}{3}-1)} + \right.\\
&~~~~~~\left.+\frac{2 P_0 \left[4 (h+5) \mu _0
   e^{\rho }-72\right] \left(1-\frac{\mu _0 e^{\rho }}{3}\right)^{h/2}}{\left(4 \mu _0
   e^{\rho }-12\right)^2}\right]\,, 
\end{align}
\end{subequations}
and
\begin{subequations}
\begin{align}
&\K=-\frac{3 K_0  e^{\rho /2}}{16 \mu _0 e^{3 \rho /2}-12 K_0 e^{\rho /2}}\,,\\
&Y=\frac{\mu _0 e^{\rho } \left(3K_0- \mu _0 e^{\rho }\right)^{h/2}}{2 \left(\mu _0 e^{\rho } -3K_0\right) \left[\left(3K_0- \mu _0 e^{\rho } \right)^{h/2}+ 9 K_0  P_0\right]}\,.
\end{align}
\end{subequations}
This solution corresponds to the  Bowers-Liang solution for a constant density object~\cite{Bowers1974abc}, which in radial coordinates reads
\begin{subequations}
\begin{align}
& d s^2 = -Ad t^2+ B d r^2 +r^2\,\big(d\theta^2+\sin^2\!\theta d\phi^2\big)\,,\\
& A =A_0 \left[\left(1-\frac{\mu _0 }{3}r^2\right)^{h/2}+3 P_0\right]^{2/h}\,,\\
& B =\frac{3 }{3 -\mu _0 r^2}\,,
\end{align}
\end{subequations}
with
\begin{subequations}
\begin{eqnarray}
&& p_r=-\frac{\mu _0 \left[\left(1-\frac{\mu _0 r^2}{3}\right)^{h/2}+P_0\right]}{\left(1-\frac{\mu _0 r^2}{3}\right)^{h/2}+3
   P_0}\,,\\
&&p_\perp =p_r-\frac{ \mu _0^2 P_0 r^2 (1-h)\left(1-\frac{\mu _0 r^2}{3}\right)^{\frac{h}{2}-1}}{3 \left[\left(1-\frac{\mu _0
   r^2}{3}\right)^{h/2}+3 P_0\right]^2}\,,
\end{eqnarray}
\end{subequations}
and $P_0<0$. 

Another interesting strategy to obtain exact solutions of Eqs.~\eqref{112TOVGen} is to find a convenient constraint for the anisotropic pressure term. For example, setting $\mathcal{P}=P+\mathbb{P}$ and imposing  
\begin{equation}\label{ConstrAniP}
\mathbb{P}=\frac{1}{6}\mathbb{M}\left(1-4\K\right)\,,
\end{equation}
the first of Eqs.~\eqref{112TOVGen} reduces to 
\begin{equation} \label{CalPEq}
\mathcal{P}_{,\rho }+\mathcal{P}^2+\mathcal{P} \left(3 \mathcal{K}-\mathbb{M}-\frac{7}{4}\right)=0\,,
\end{equation}
which is analogous to the TOV equation the pressure of an isotropic system with pressure $\mathcal{P}$. It is easy to check that in the trivial case $\mathcal{P}=0$, the above equations give the solution found by Florides~\cite{Florides1974abc}. As a consequence, the Florides' solution  can be considered the simplest  element of an entire class of solutions for which Eq.~\rf{ConstrAniP} holds. 

Using the covariant TOV equations one can easily find some other simple examples with a non trivial $\mathcal{P}$. Indeed, working with a constant density, we have
\begin{subequations}
\begin{align}
& d s^2 = -Ad t^2+ B d r^2 +r^2\,\big(d\theta^2+\sin^2\!\theta d\phi^2\big)\,,\\
& A =\frac{A_0
   \left(e^{\mathcal{P}_0}r_0^2+r^2\right)^2}{\sqrt{3 - \mu
   _0 r^2}}\,,\\
& B =\frac{3}{3- \mu _0 r^2}\,,
\end{align}
\end{subequations}
together with the radial and tangential pressure 
\begin{subequations}
\begin{eqnarray}
&& p_r= \frac{3 - \mu _0 r^2}{3 \left(r^2+ e^{\mathcal{P}_0}r_0^2\right)}\,,\\
&&p_\perp =\frac{\mu _0^2 r^2}{4(3 - \mu _0 r^2)}+p_r\,.
\end{eqnarray}
\end{subequations}
The solution is well behaved in the origin. Unfortunately in the value of $r$ in which the radial pressure becomes zero the orthogonal pressure diverges, so the solution can only be used  by choosing matching it  to the Schwarzschild exterior before this singularity. Naturally, since $p_r\neq0$ at the junction  the object represented by this solution will be enclosed in a shell, defined by the stress-energy tensor in Eq.~\eqref{IsraelShell}.  

A more regular example is (we set $r_0 =1$ for brevity):
\begin{subequations}
\begin{align}\label{ABGenFlor}
& d s^2 = -Ad t^2+ B d r^2 +r^2\,\big(d\theta^2+\sin^2\!\theta d\phi^2\big)\,,\\
& A =A_0 \psi  \left(12 a^2-3 a (4 b-1) r^2+b (4 b-2)
   r^4\right)^{\frac{4 b-5}{8 b-4}}\,,\\
& \psi=\exp \left(\frac{\sqrt{3} (4 b+1) \tan ^{-1}\left(\frac{a (3-12 b)+2 b (4 b-2) r^2}{a
   \sqrt{48 b^2-24 b-9}}\right)}{(2-4 b) \sqrt{(4 b-3) (4 b+1)}}\right)\,,
\\
& B =\frac{12 \left(b r^2-a\right)^2}{12
   a^2-3 a (4 b-1) r^2+b (4 b-2) r^4}\,,
\end{align}
\end{subequations}
where $a$ is an arbitrary constant and $b<-1/4$ or $b>3/4$. The thermodynamical quantities are: 
\begin{subequations}\label{MPPPGenFlor}
\begin{eqnarray}
&& \mu=\frac{(4 b+1) \left(9 a^2-7 a b r^2+2 b^2 r^4\right)}{12 \left(b
   r^2-a\right)^3}\,,\\
&& p_r=\frac{1}{a-b r^2}\,,\\
&&\nn p_\perp =\frac{\left[4 b^2
   r^4+3 a \left(4 a+r^2\right)-b \left(12 a r^2+2 r^4\right)\right]^{-1}}{48 \left(b
   r^2-a\right)^3 }\times\\
&&~~~~~\nn\left\{9 a^3 \left(64 a+19 r^2\right)\right.\\
&&~~~~~ \nn-b^3 \left(624 a^2 r^4+800
   a r^6+100 r^8\right)\\
&&~~~~~\nn+a b^2 r^2 \left(432 a^2+1608 a r^2+356 r^4\right)\\
&&\nn~~~~~-9 a^2 b r^2
   \left(168 a+47 r^2\right)\\
   && ~~~~~\left.+80 b^4 r^6 \left(4 a+2 r^2\right)-64 b^5 r^8\right\}.
\end{eqnarray}
\end{subequations}
Notice that the radial pressure in this solution is never zero. Hence, in principle,  we have  two different options. A first one is   to consider the metric above as representing an object with a thin atmosphere that covers the entire spacetime. A second option would be to  make a junction with the Schwarzschild solution by introduction a thin shell, like in the previous example. However, the component $A$ of the metric and the barotropic factors are growing functions of the radial coordinate. This implies that the first option would lead to an unphysical situation and suggests a natural range of radii in which the solution can be soldered to the Schwarzschild solution. 

\begin{figure}[!ht]
    \subfloat[The coefficients of the metric \rf{ABGenFlor}. The blue line represents $A$ and the orange $B$. \label{F:ABGenFlor}]{%
      \includegraphics[width=0.45\textwidth]{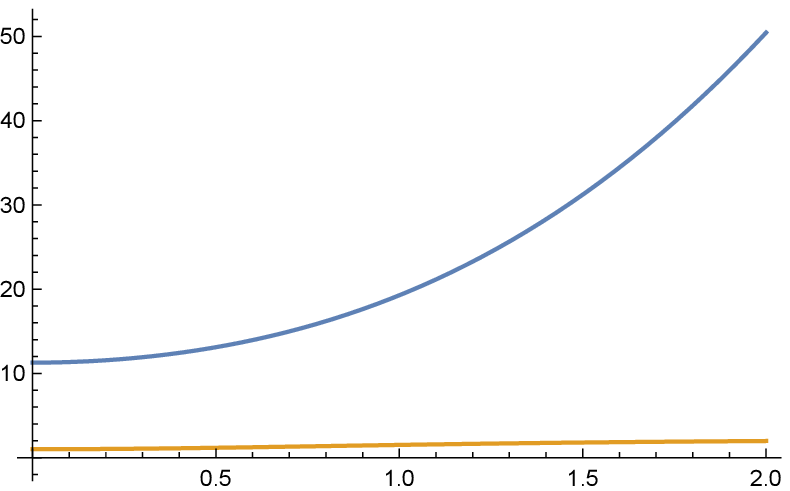}}\\\mbox{}\\
    \subfloat[The thermodynamic quantities \rf{MPPPGenFlor}  associated with \rf{ABGenFlor}. The blue line represents  $p_r$, the orange $p_\perp$ and the green $\mu$.\label{F:MPPPGenFlor}]{%
      \includegraphics[width=0.45\textwidth]{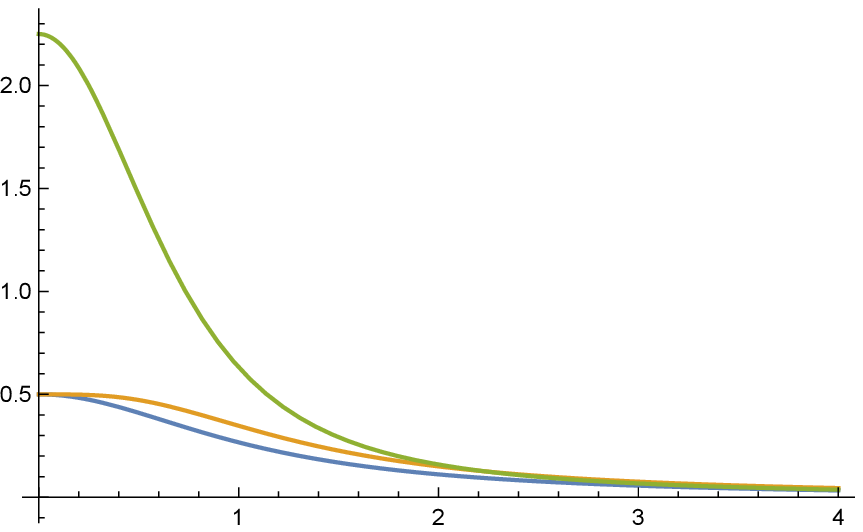}}\\
      \subfloat[The equation of state  associated with \rf{ABGenFlor}.  \label{F:wGenFlor}]{%
      \includegraphics[width=0.45\textwidth]{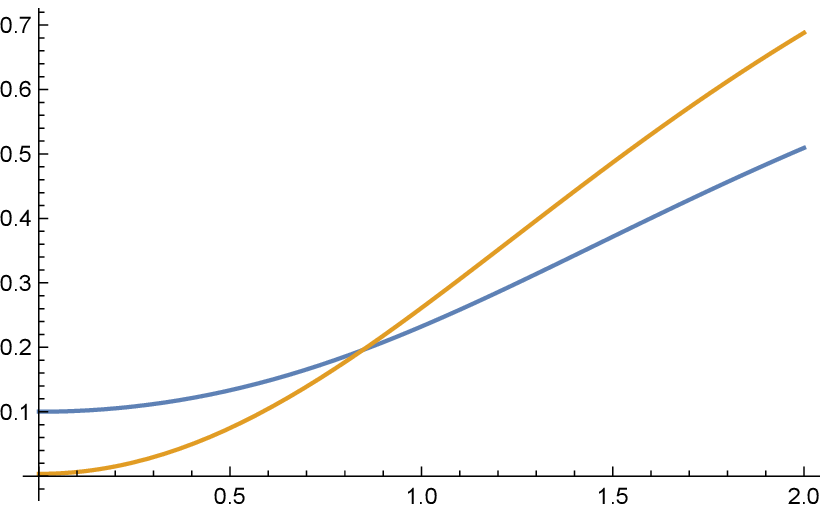}}
      \caption{Graphs of the solution \rf{ABGenFlor} in the case $r_0=1$,  $a=2$, $\mu_1=-1$, $a=-7/4$. The values of the parameters has been chosen in such a way to make the features of the solution as clear as possible.}
          \label{fig:GenFlorides}
  \end{figure}

In Fig.~\ref{fig:GenFlorides} we give the behaviour of the solution for a convenient choice of the parameters.

Yet a different resolution strategy for Eqs.~\eqref{112TOVGen} consists in separating the isotropic and anisotropic degrees of freedom. Shifting the position of the coupling term $2 P \mathbb{P}$ and of the term  $\left(\frac{1}{4}-\mathcal{K}\right) \mathbb{M}$  one can decompose this equation in a system of equations which, given a choice for the behaviour of the energy density, is composed by a Bernoulli and a Riccati equation.

A particularly interesting way to perform this separation is: 
\begin{equation}\label{PTOVAnis}
\begin{split}
&P_{,\rho }+P^2-P \left(\mathbb{M}-3 \mathcal{K}+\frac{7}{4}\right)\\
&~~~~~~~~~~~~~+\left(\frac{1}{4}-\mathcal{K}\right) \mathbb{M}=\\
&~~~~~~~~~~~~~\mathbb{P}_{,\rho }-\mathbb{P}^2+\mathbb{P} \left(2P+\mathbb{M}-3
   \mathcal{K}+\frac{1}{4}\right),\\
&\mathcal{K}_{,\rho }=-2 \mathcal{K}\left(\mathcal{K}-\frac{1}{4}-\mathbb{M}\right).
\end{split}
\end{equation}
This setting suggests that any isotropic ($\mathbb{P}=0$) solution of the TOV equations can be associated to an anisotropic solution in which $P$, $\K$ and $\mathbb{M}$ are the same and the $\mathbb{P}$  is determined by the equation
\begin{equation}
\mathbb{P}_{,\rho }-\mathbb{P}^2+\mathbb{P} \left(2P+\mathbb{M}-3
   \mathcal{K}+\frac{1}{4}\right)=0\,.
\end{equation}
Notice, however, that these new anisotropic solutions will have a different $Y$ and therefore a different $(0,0)$ component of the metric because the constraint \rf{NewConstrGenV2} is changed by the presence of $\mathds{P}$.  

\begin{figure}[!ht]
    \subfloat[The coefficients of the metric \rf{NewSolAni-1}. The blue line represents $A$ and the orange $B$. \label{F:NewSolAni-1}]{%
      \includegraphics[width=0.45\textwidth]{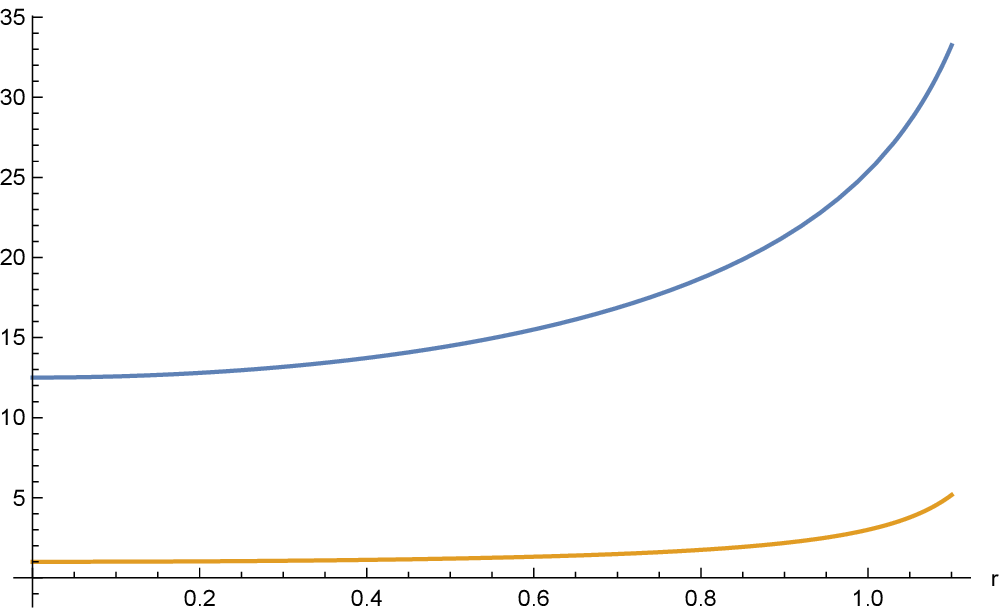}}\\\mbox{}\\
    \subfloat[The thermodynamic quantities \rf{NewSolAni-2}  associated with \rf{NewSolAni-1}. The blue line represents  $p_r$, the orange $p_\perp$ and the green $\mu$.\label{F:NewSolAni-2}]{%
      \includegraphics[width=0.45\textwidth]{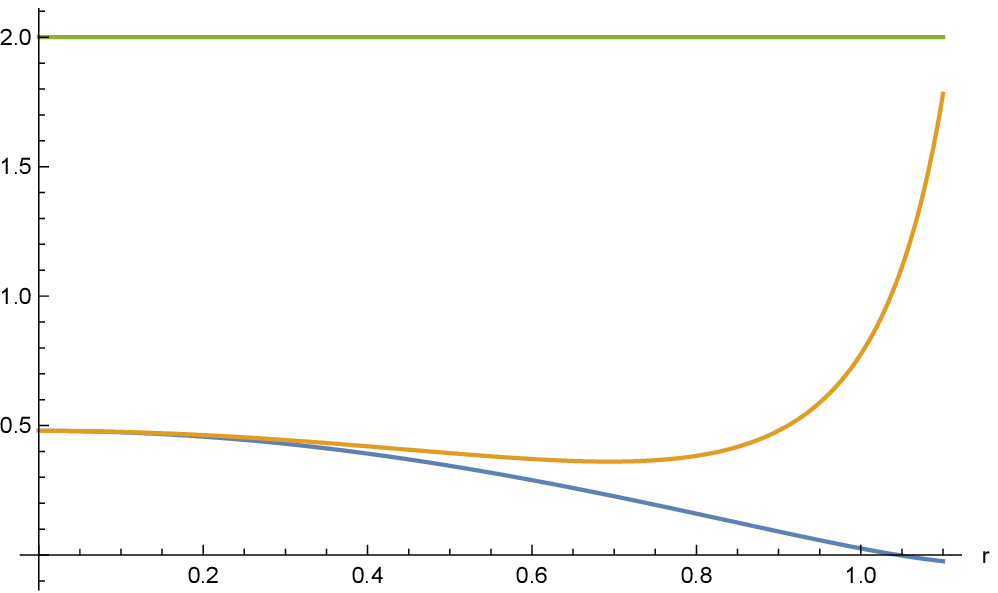}}
      \caption{Graphs of the solution \rf{NewSolAni-1} in the case $r_0=1$, $\mu_0=3/2$, $\mu_1=-1/2$,  $\beta=-1$, $P_*=-7$ . The values of the parameters have been chosen in such a way to make the features of the solution as clear as possible.}
          \label{F:NewSolAni}
  \end{figure}

One can indeed find a number of different ways to connect isotropic and anisotropic solutions.
For example, setting 
\begin{equation}
P=P_0-\mathbb{P}, \qquad \mathbb{M}=\mathbb{M}_0+\alpha \mathbb{M}_1\,,
\end{equation}
where $P_0$ and  $\mathbb{M}_0$ are part of a known solution of the isotropic equations. With these assumptions the Eqs.~\rf{PTOVAnis} returns an algebraic equation that can be solved for $\mathbb{P}$:
\begin{equation}
\begin{split}
\mathbb{P}=&\frac{1}{6} \left[P_0 \left(4 \alpha  \mathbb{M}_1+4
   \mathbb{M}_0-12 \mathcal{K}+7\right)-4 P_{0, \rho} \right.\\
 &\left.-4 P_0^2-(4 \mathcal{K}-1) \left(\alpha
   \mathbb{M}_1+\mathbb{M}_0\right)\right]\,.
\end{split}
\end{equation}
Now the TOV equations can be solved completely if one is able to integrate the second of Eqs.~\rf{PTOVAnis}. The latter equation is of Bernoulli type and admits the following formal solution:
\begin{equation} \label{SolKAniStr}
\begin{split}
&\K=\frac{e^{F}}{ \K_{*} -2 \int e^{F} d\rho}\,, \\
&F=  \int \frac{1}{2} \left(4 \alpha  \mathbb{M}_1+4 \mathbb{M}_0+1\right)d\rho\,.
\end{split}
\end{equation}
As an example, one can start with the classical isotropic constant density  solution
\begin{subequations}
\begin{align}\label{MuConstFl}
& d s^2 = -Ad t^2+ B d r^2 +r^2\,\big(d\theta^2+\sin^2\!\theta d\phi^2\big)\,,\\
& A =A_0\left(\sqrt{3 - \mu _0 r^2}+P_0\right)^2\,, \label{MuConstFlA}\\
& B =\frac{3}{3 -\mu _0 r^2}\,,
\end{align}
\end{subequations}
with
\begin{equation}
p=-\frac{\mu _0\left(P_0+3 \sqrt{3-\mu _0 r^2}\right)}{3 \left(P_0+\sqrt{3-\mu _0
   r^2}\right)}\,.
\end{equation}
Setting
\begin{equation}
\mathbb{M}_1=\frac{ \mu_1 e^{\rho }}{K_0} \mathcal{K}\,, 
\end{equation}
where $\mu_1$ is a constant, the solution for $\mathcal{K}$ is
\begin{equation} 
\begin{split}
\mathcal{K}=\frac{3 K_0}{4 \left(3 K_0-4 e^{\rho } \left(\alpha  \mu _1+\mu _0\right)\right)}\,.
\end{split}
\end{equation}
This leads, after long but trivial calculations, to the following solution for the metric and the fluid thermodynamics: 
\begin{subequations}\label{NewSolAni-1}
\begin{align}
& d s^2 = -A d t^2+ B d r^2 +r^2\,\big(d\theta^2+\sin^2\!\theta d\phi^2\big)\,,\\
& A =\frac{A_0  \sqrt{3-\mu _0 r^2}
   \left(3-P_0^2r_0^2-\mu _0 r^2\right) }{\sqrt{3- r^2 \left(\alpha  \mu _1+\mu _0\right)}}\times\\
&\exp \left(2 \tanh ^{-1}\left(\frac{\sqrt{ 3-\mu _0 r^2}}{P_0 r _0}\right)\right)\,,\\
& B =\frac{3}{3 - r^2 \left(\alpha  \mu _1+\mu _0\right)}\,,
\end{align}
\end{subequations}
where $P_0$ is an integration constant. The radial and tangential pressure read
\begin{subequations}\label{NewSolAni-2}
\begin{align}
& p_r=\frac{\mu _0  \left[3-
   r^2 \left(\alpha  \mu _1+\mu _0\right)\right]\left(3 \sqrt{3 - \mu _0
   r^2}+P_0r_0\right)}{3\left( \mu _0 r^2-3 \right) \left(\sqrt{3 - \mu _0 r^2}+P_0r_0\right)}\,,\\
& \nn p_\perp =-192 \left(\mu _0 r^2-3\right){}^2 \left(r^2 \left(\beta  \mu _1+\mu
   _0\right)-3\right) p_r\\ & 
   -\frac{36 \beta  \mu _1 r^2 \left(8
   \mu _0 \left(r^2 \left(\beta  \mu _1+\mu _0\right)-3\right)-12 \beta  \mu
   _1\right)}{48 \left(\mu _0 r^2-3\right){}^2 \left(12-4 r^2 \left(\beta  \mu _1+\mu
   _0\right)\right)}\,.
\end{align}
\end{subequations}
In this solution the pressures at the centre are regular for any value of the parameters and one can have central value of these quantity to be positive. In addition, the parameters can be set in such a way to avoid any singularity. Fig.~\ref{F:NewSolAni} gives an example in which the radial pressure goes to zero at a finite radius. 

The reason behind the connection between isotropic and anisotropic solutions will become clear in the next Section when we will look into the details of the generating theorems for anisotropic solutions. We will discover that ultimately these theorems are behind the methods presented above. 

\section{Generating theorems for anisotropic TOV solutions}
In Refs.~\cite{Boonserm:2005ni,Boonserm:2006up,ExactSolutions} a number of interesting theorems were proved, dubbed {\it generating theorems}. These theorems allow to connect different solutions of the TOV equations in the isotropic case, in the sense that given a solution of the equations, one can recover new solutions which differ only in a given set of quantities ({\it e.g.} the pressure and $(0,0)$ component of the metric). In Ref.~\cite{SDTOV1}  using the $1+1+2$ covariant formalism, we were able to prove that these theorems can be related to linear deformations of the solutions of the isotropic TOV equations. We also showed that the constraints between the variables $Y, \K, P, \M$ can be used  as a guide to deduce such generating theorems. In this Section we will extend these generating theorems to the case of anisotropic solutions. We will show that, in addition to the theorems already found for the isotropic case, new theorems can be formulated. 

Let us start from  a shift in the isotropic pressure and the variable $Y$:
\begin{equation}\label{TeoGen1}
\begin{split}
&P= P_0 + P_1\,,\qquad Y= Y_0 + Y_1\,.
\end{split}
\end{equation}
which corresponds to the transformation
\begin{subequations}
\begin{align}
& A \rightarrow A_0 (\rho) \exp\left(\int Y_1 d\rho\right)~,\\
& B \rightarrow B_0 (\rho)~,\\
& C\rightarrow C_0 (\rho)~.
\end{align}
\end{subequations}
of theorem 2 in Ref.~\cite{Boonserm:2005ni}.

Using the Eqs.~\rf{PTOVAnis}, the constraint in Eq.~\rf{NewConstrGenV2} and Eq.~\rf{NewEqGenV1}, we obtain
\begin{equation}\label{TeoGen2Eq}
\begin{split}
&P_{1, \rho}+P_1^2+P_1 \left(3 \mathcal{K}_0-\mathbb{M}_0+2 P_0+2\mathbb{P}_0-\frac{7}{4}\right)=0~, \\
& Y_1=P_1~.
\end{split}
\end{equation}
whose solution is
\begin{equation}\label{TeoGen2}
\begin{split}
& P_1=\frac{e^{F}}{ P_{*} + \int e^{F} d\rho}\,,\\
&F= \int \left(3 \mathcal{K}_0-\mathbb{M}_0+2 P_0+2\mathbb{P}_0-\frac{7}{4}\right) d\rho\,.
\end{split}
\end{equation}
In other words starting form the solution $(Y_0,\mathcal{K}_0,P_0,\mathbb{M}_0)$ we have obtained  a new solution $(Y,\mathcal{K},P,\mathbb{M})$ by solving two integrals. 

In~\cite{SDTOV1} we proved a similar theorem for the isotropic case. Comparing the two results it is easy to see that the only difference consists in the presence of the additional term $2\PP$ in the integral that defines $F$.
 
Let us consider now the case of a combined shift of the pressure and the energy density. Setting
\begin{equation}\label{TeoGen2}
\begin{split}
&P= P_0 + P_1\,,\qquad \M= \M_0 + \M_1\,,\\
&\K= \frac{1}{\K_0+\K_1}\,,
\end{split}
\end{equation}
which leads to 
 \begin{subequations}
\begin{align}
& A \rightarrow A_0 (\rho)\,,\\
& B^{-1} \rightarrow B_0 (\rho) + \frac{ e^\rho}{K_0}\K_1\,,\\
& C\rightarrow C_0 (\rho)\,,
\end{align}
\end{subequations}
and corresponds to Theorem 1 in Ref.~\cite{Boonserm:2005ni} (see also~\cite{ExactSolutions}). Eq.~\rf{PTOVAnis}, the constraint in Eq.~\rf{NewConstrGenV2} and Eq.~\rf{NewEqGenV1} return 
\begin{equation}
\begin{split}
&\mathcal{K}_{1,\rho}=-\mathcal{K}_1 \Phi+\Gamma\,,\\
& \Phi=\frac{12 \mathcal{K}_0-4 \mathbb{M}_0-1+Y_0 \left(8 \mathcal{K}_0-8 \mathbb{M}_0-2\right)}{ 2(1+2 Y_0)}\,,\\
& \Gamma= \frac{8 \mathcal{K}_0^2\left(Y_0+1\right)+ \mathcal{K}_0 \left(4 \mathbb{M}_0+1\right) \left(2
   Y_0+1\right)-4}{ 1+2 Y_0}\,,\\
& P_1=\frac{\mathcal{K}_0^2+\mathcal{K}_1 \mathcal{K}_0-1}{\mathcal{K}_0+\mathcal{K}_1}\,,\\
&\M_1=-\frac{\left[\mathcal{K}_0 \left(\mathcal{K}_0+\mathcal{K}_1\right)-1\right] \left(2
   Y_0+3\right)}{\left(\mathcal{K}_0+\mathcal{K}_1\right) \left(2 Y_0+1\right)}\,.
\end{split}
\end{equation}
The first equation above is a linear differential equation which can always be solved exactly as
\begin{equation}
\begin{split}
&\K_1=e^{-F} \left(\K_{*} -\int e^{F} \Gamma d\rho\right)\,, \\
&F= \int \Phi d\rho\,,
\end{split}
\end{equation}
Differently from the previous case, the theorem that we have obtained matches exactly the corresponding theorem in the isotropic case given in~\cite{SDTOV1}.  

Exploiting the similarity of the role of the isotropic and anisotropic pressures in Eqs.~\eqref{112TOVGen}, we can, moreover, give additional theorems. For example, keeping the isotropic pressure unchanged one can set:
\begin{equation}\label{TeoGen1PP}
\begin{split}
&\PP= \PP_0 + \PP_1\,,\qquad Y= Y_0 + Y_1\,,
\end{split}
\end{equation}
and we obtain
\begin{equation}\label{TeoGenPP1}
\begin{split}
& Y_1=\PP_1\,,\\
& \PP_1=\frac{e^{F}}{ \PP_{*} + \int e^{F} d\rho}\,,\\
&F=\int \left(3 \mathcal{K}_0-\mathbb{M}_0+2 P_0+2 \PP_0-\frac{1}{4}\right) d\rho\,,
\end{split}
\end{equation}  
which mirrors theorem in Eq.~\eqref{TeoGen1}.

Also theorem in Eq.~\eqref{TeoGen2} has a similar analogue. Indeed, setting
\begin{equation}\label{TeoGen2PP}
\begin{split}
&\PP= \PP_0 + \PP_1\,,\qquad \M= \M_0 + \M_1\,,\\
&\K= \frac{1}{\K_0+\K_1}\,,
\end{split}
\end{equation}
we obtain
\begin{equation}
\begin{split}
&\mathcal{K}_{1,\rho}=\mathcal{K}_1 \Phi+\Gamma\,,\\
& \Phi=\frac{Y_0 \left(8 \mathcal{K}_0-8 \mathbb{M}_0-2\right)-4 \mathbb{M}_0-1}{2(1+2 Y_0)}\,,\\
& \Gamma= \frac{\mathcal{K}_0^2 \left(8
   Y_0+2\right)-\mathcal{K}_0 \left(4 \mathbb{M}_0+1\right) \left(2 Y_0+1\right)+2}{1+2 Y_0}\,,\\
& \PP_1=\frac{\mathcal{K}_0^2+\mathcal{K}_1 \mathcal{K}_0-1}{\mathcal{K}_0+\mathcal{K}_1}\,,\\
& \M_1=-\frac{2 \left[\mathcal{K}_0 \left(\mathcal{K}_0+\mathcal{K}_1\right)-1\right]
   Y_0}{\left(\mathcal{K}_0+\mathcal{K}_1\right) \left(2 Y_0+1\right)}\,.
\end{split}
\end{equation}
Again, the first equation above is a linear differential equation which can be solved exactly as:
\begin{equation}
\begin{split}
&\K_1=e^{F} \left(\K_{*} -\int e^{-F} \Gamma d\rho\right)\,, \\
&F= \int \Phi d\rho\,.
\end{split}
\end{equation}
A third generating theorem allows to shift the isotropic pressure, maintaining the energy density fixed. Setting
\begin{equation}\label{TeoGen2_3}
\begin{split}
&\PP= \PP_0 + \PP_1,\qquad P= P_0 + P_1\,,\\
&\K= \frac{1}{\K_0+\K_1}\,,
\end{split}
\end{equation}
we obtain
\begin{equation}
\begin{split}
&\mathcal{K}_{1,\rho}=-\mathcal{K}_1 \Phi+\Gamma\,,\\
& \Phi=\frac{1}{2}\left(4 \mathbb{M}_0+1\right)\,,\\
&\Gamma=\frac{1}{2}\left[ 4 \mathcal{K}_0^2-2 \mathcal{K}_0 \left(4 \mathbb{M}_0+1\right)+4\right]\,,\\
& \PP_1=\frac{\left(\mathcal{K}_0 \left(\mathcal{K}_0+\mathcal{K}_1\right)-1\right) \left(2
   Y_0+3\right)}{3 \left(\mathcal{K}_0+\mathcal{K}_1\right)}\,,\\
& P_1=\frac{2}{3} \left(\frac{1}{\mathcal{K}_0+\mathcal{K}_1}-\mathcal{K}_0\right) Y_0\,,
\end{split}
\end{equation}
which can be solved by
\begin{equation}\label{TheoKDiffEq}
\begin{split}
&\K_1=e^{-F} \left(\K_{*} -\int e^{F} \Gamma d\rho\right)\,, \\
&F= \int \Phi d\rho\,.
\end{split}
\end{equation}
Like in the isotropic case, one can further consider nonlinear deformations of a known solution.  For example, in the case of a linear shift of the isotropic and anisotropic pressure with a generic change of $\K$ that is
\begin{equation}\label{TeoGen2_3}
\begin{split}
&\PP= \PP_0 + \PP_1\,,\qquad P= P_0 + P_1\,,\\
&\K= \K_1\,,
\end{split}
\end{equation}
we obtain
\begin{equation}
\begin{split}
&\mathcal{K}_{1,\rho}=-\mathcal{K}_1 \left(-2 \mathbb{M}_0-\frac{1}{2}\right)-2 \mathcal{K}_1^2\,,\\
& \PP_1=\left(2 \mathcal{K}_0-1\right)
   P_0-\mathbb{P}_0+Y_0+\frac{1}{4}\\
&~~~~~\frac{1}{6}
   \mathcal{K}_1 \left(-4 P_0-4 \mathbb{P}_0-5\right)+\frac{2}{3} \mathcal{K}_0 \left(\mathbb{M}_0+3 \mathbb{P}_0\right)\,,\\
& P_1=\frac{1}{6} \mathcal{K}_1 \left(4 P_0+4 \mathbb{P}_0-1\right)\\
&~~~~~~~-\frac{2}{3} \mathcal{K}_0
   \left(\mathbb{M}_0+3 P_0+3 \mathbb{P}_0\right)\,.
\end{split}
\end{equation}
Similar results can also be found if one considers variations of $P$ and $\M$ or $\PP$ and $\M$ together with a change of $\K$. As in the isotropic case, one can obtain hints about the existence of new theorems by looking at the constraints in Eqs.~(\ref{NewConstrGenV1})-(\ref{NewConstrGenV3}).

\section{Reconstructing anisotropic solutions}

To conclude, we give a reconstruction algorithm for  anisotropic stellar interior solutions. In Ref.~\cite{SDTOV1} we proposed a similar algorithm for isotropic  stellar interior solutions. In that case it turned out that the metric coefficients have to satisfy a differential constraint which can be difficult to solve. We will make a similar construction here, showing that, surprisingly, the algorithm for anisotropic solutions does not present any such constraints. 

Solving Eqs.~(\ref{NewEqGenV1})-(\ref{NewConstrGenV3}) for the matter variables, one obtains:
\begin{align}
&\mathbb{M}=\frac{\mathcal{K}_{\rho}}{2 \mathcal{K}}+\mathcal{K}-\frac{1}{4}\,,\label{ANI_Rec1}\\
&\nn P=
   \frac{1}{3} \left(2 Y_{\rho}+2 Y^2+ Y\right)\\
&~~~~~~-\frac{2 Y+1}{6} \frac{
   \mathcal{K}_{\rho}}{\mathcal{K}}-\frac{1}{3} \mathcal{K}+\frac{1}{12}\,,\\
&\nn \PP=(2 Y+1) \mathcal{K}_{\rho}-4 \mathcal{K}^2\\
&~~~~~~-\mathcal{K} \left[4
   Y_{\rho}+4 (Y-1) Y-1\right]\,. \label{ANI_Rec3}
\end{align}
In the isotropic case one should set $\PP=0$ in \eqref{ANI_Rec3} and this relation corresponds to the differential constraint we encountered in Ref.~\cite{SDTOV1}. This result implies that the differential constraint found in the isotropic case corresponds to the very condition of isotropy.  

The structure of Eqs. (\ref{ANI_Rec1}-\ref{ANI_Rec3}) shows  that reconstructing anisotropic solutions is considerably simpler than reconstructing isotropic ones. Indeed in the anisotropic case the equation above leads to a double infinity of solutions.

It is useful at this stage to give the expression of the barotropic factors in terms of the variables $\K$ and $Y$:
\begin{align}
&w_r=\frac{ \mathcal{K}^2\left(-4 \mathcal{K}-4 Y_{,\rho }+4 Y+1\right)+(4 Y+1) \mathcal{K}\mathcal{K}_{,\rho }}{4
   \mathcal{K}^3-\mathcal{K}^2+\mathcal{K} \left(\mathcal{K}_{,\rho }-2 \mathcal{K}_{,\rho \rho }\right)+4 \mathcal{K}_{,\rho }^2}\,,\\
   &w_\perp =-\frac{1+16
   \mathcal{K} Y^2 \left(\mathcal{K}_{,\rho }+\mathcal{K}\right)+\mathcal{K}\mathcal{K}_{,\rho } \left(1-12 Y_{,\rho }\right) }{\left[4
   \mathcal{K}^3-\mathcal{K}^2+\mathcal{K} \left(\mathcal{K}_{,\rho }-2 \mathcal{K}_{,\rho \rho }\right)+4 \mathcal{K}_{,\rho }^2\right]}\\
   \nn&~~~~~~+\frac{ 2\mathcal{K}^2 \left(-8 Y_{,\rho }+8
   Y_{,\rho \rho }+1\right)-8 \mathcal{K}^3}{\left[4
   \mathcal{K}^3-\mathcal{K}^2+\mathcal{K} \left(\mathcal{K}_{,\rho }-2 \mathcal{K}_{,\rho \rho }\right)+4 \mathcal{K}_{,\rho }^2\right]}\\
    &~~~~~~+\frac{ Y \left(4 \mathcal{K} \left(\mathcal{K}_{,\rho
   }+\mathcal{K} \left(-4 \mathcal{K}+8 Y_{,\rho }+1\right)\right)-2\right)}{\left[4
   \mathcal{K}^3-\mathcal{K}^2+\mathcal{K} \left(\mathcal{K}_{,\rho }-2 \mathcal{K}_{,\rho \rho }\right)+4 \mathcal{K}_{,\rho }^2\right]}\,.
\end{align}
These expressions can be used to select the suitable forms of the metric variables $Y$ and $\K$.

We will now use the above algorithm to generate some physically relevant solutions in the sense of Section~\ref{ANISOTOV}.  We shall limit ourselves to give all the results directly in the parameter $r$.

\begin{figure}[!ht]
    \subfloat[A semilogarithmic plot of the coefficients of the metric \rf{RecANIMetric1}. The blue line represents $A$ and the orange $B$. \label{F:RecANIMetric1}]{%
      \includegraphics[width=0.45\textwidth]{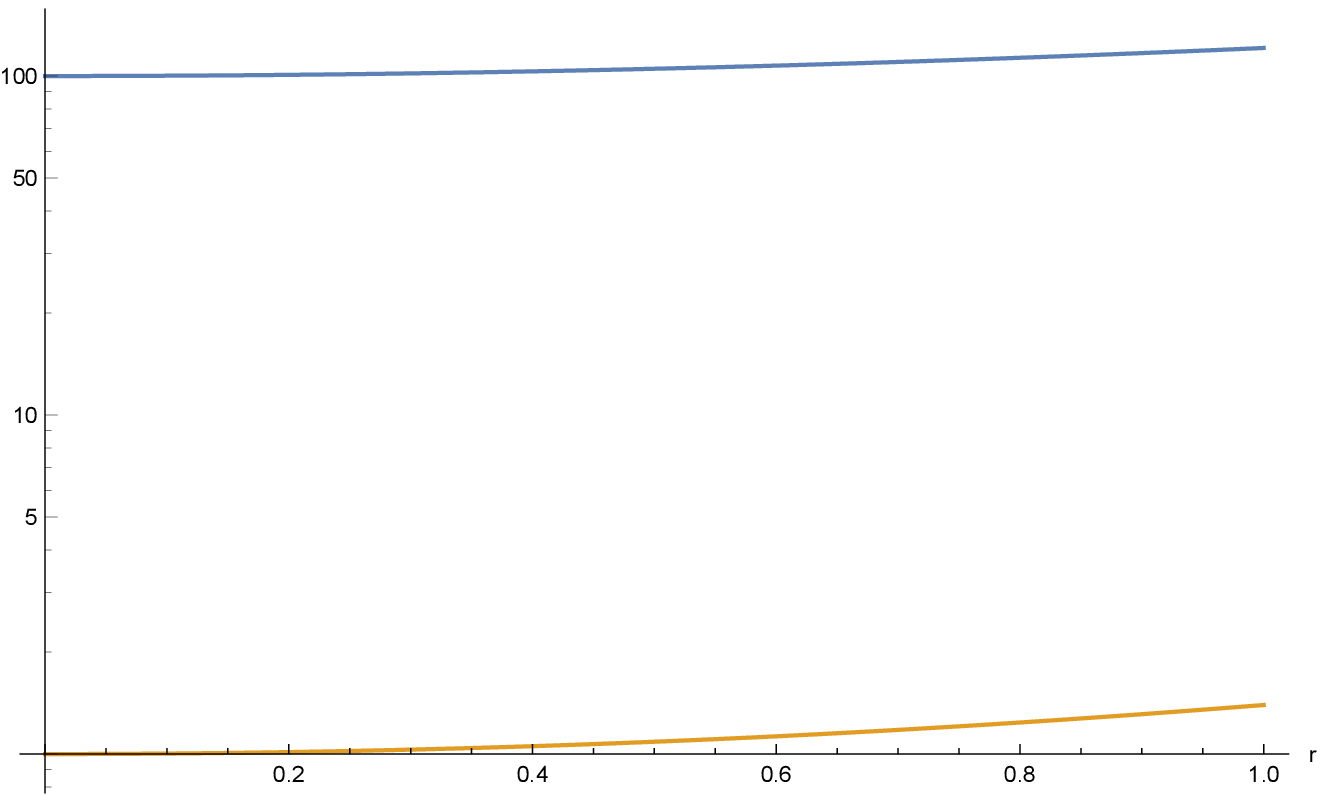}}\\
    \subfloat[A semilogarithmic plot of the thermodynamic quantities \rf{RecAni1MuPP}  associated with \rf{RecANIMetric1}. The blue line represents  $p_r$, the orange $p_\perp$ and the green $\mu$.\label{F:RecAni1MuPP}]{%
      \includegraphics[width=0.45\textwidth]{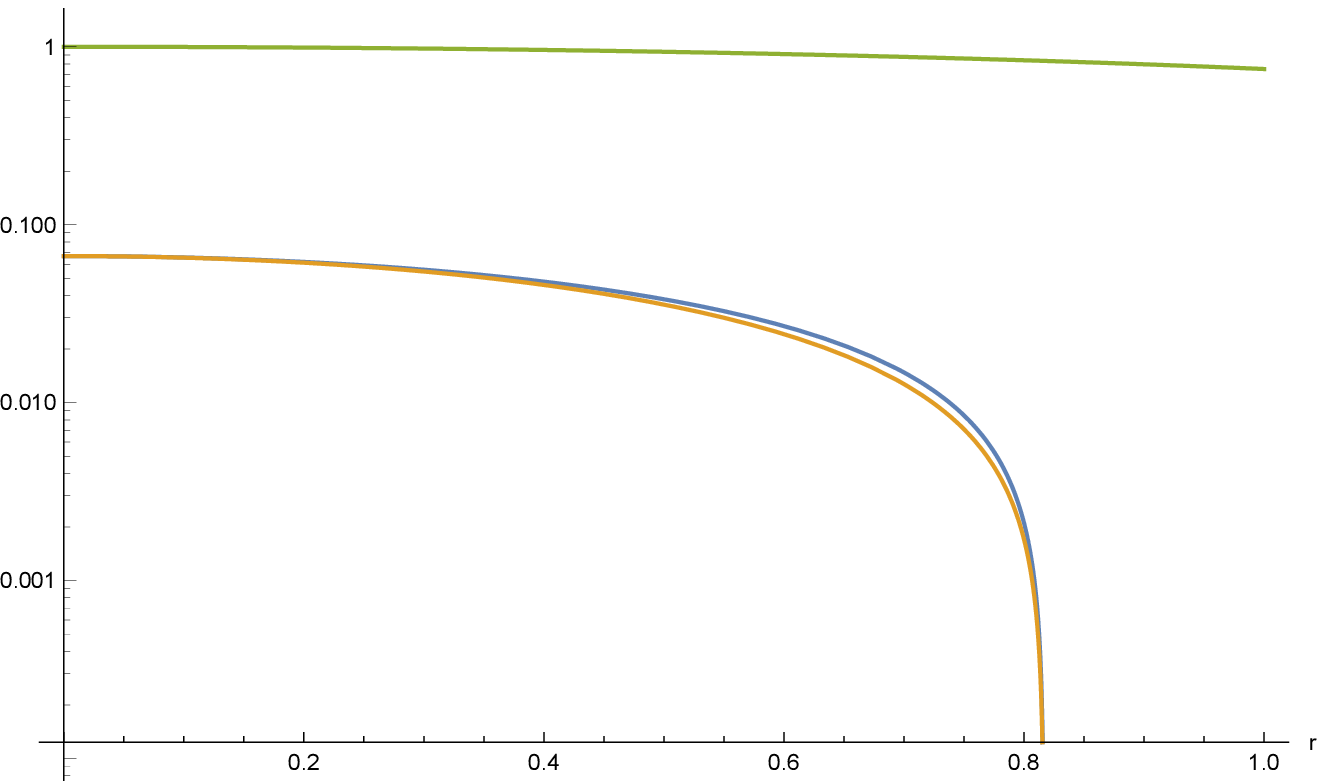}}\\
       \caption{Graphs of the solution \rf{RecANIMetric1} in the case $r_0=1$,  $\alpha=1$, $\beta=1$, $\delta=2$, $a=10$, $b=1$,  $\mu_0=1$, $\mu_1=1/4$. The values of the parameters have been chosen in such a way to make the features of the solution as clear as possible.}
          \label{F:RecANI1a}
  \end{figure}

\begin{figure}[!ht]
    \subfloat[The behaviour of the anisotropic pressure $\Pi$  associated with \rf{RecANIMetric1}.\label{F:RecAni1Pi}]{%
      \includegraphics[width=0.45\textwidth]{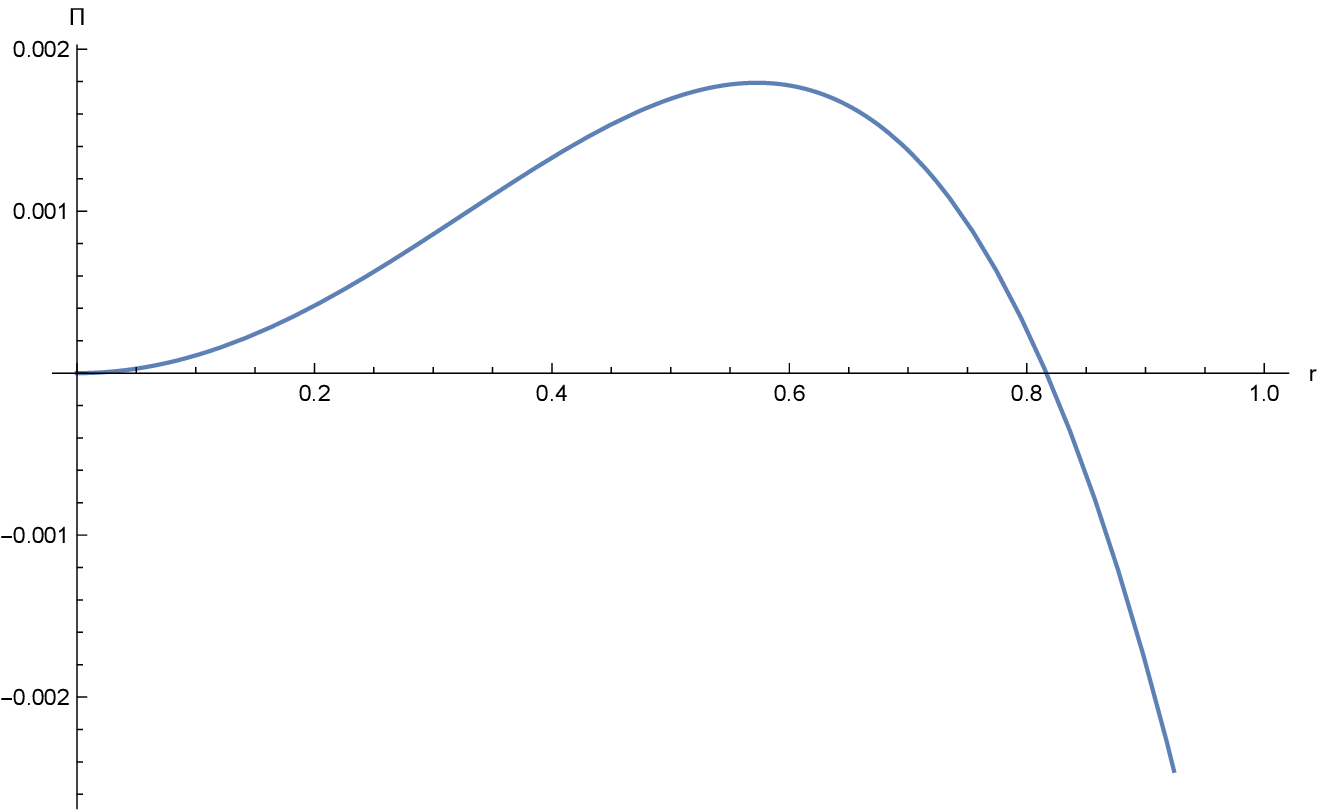}}\\
    \subfloat[The radial and orthogonal barotropic factors\rf{RecAni1wr} and \rf{RecAni1worth} (blue and orange line, respectively)   associated with \rf{RecANIMetric1}.\label{F:RecAni1ww}]{%
      \includegraphics[width=0.45\textwidth]{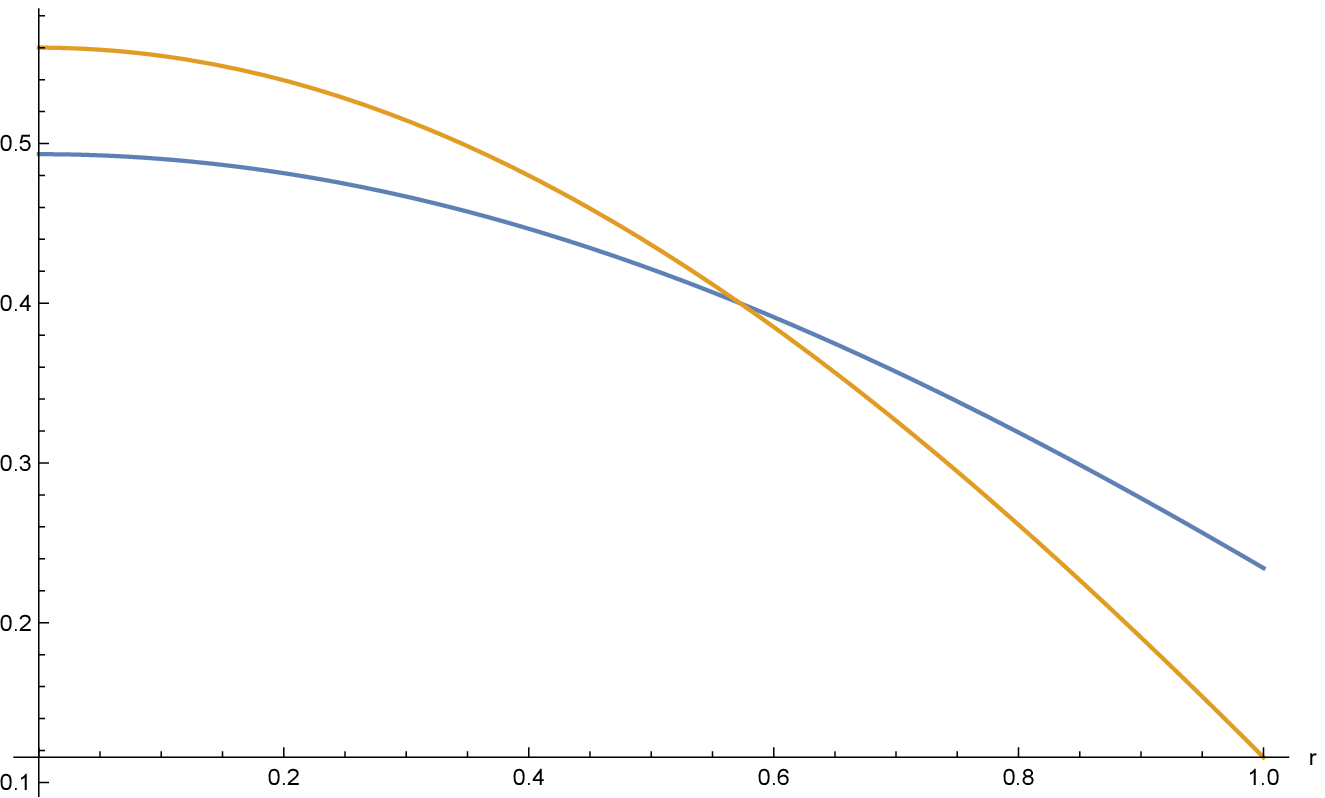}}\\
       \caption{Graphs of the solution \rf{RecANIMetric1} in the case $r_0=1$,   $\alpha=1$, $\beta=1$, $\delta=2$, $a=10$, $b=1$,  $\mu_0=1$, $\mu_1=1/4$. The values of the parameters have been chosen in such a way to make the features of the solution as clear as possible.}
          \label{F:RecANI1b}
  \end{figure}

Let us start form the metric coefficients 
\begin{equation}
\begin{split}\label{RecANIMetric1}
&A=\left(a+b r^{2\rho}\right)^\delta\,,\\
&B=\frac{3
   \mu _0}{3 \mu
   _0-r^2 \left(\mu _0-\mu _1 r^{2 \alpha }\right){}^{\beta +1}\mathcal{F}}\,,
   \end{split}
   \end{equation}
where $\mathcal{F}$ is the Gaussian hypergeometric function
\begin{equation}
\mathcal{F}=\,
   _2F_1\left(1,\beta +\frac{3}{2 \alpha }+1;1+\frac{3}{2 \alpha };\frac{r^{2\alpha  \rho
   } \mu _1}{\mu _0}\right)\,.
\end{equation}
In this way, setting for brevity $r_0=1$, we get:
\begin{equation}\label{RecAni1MuPP}
\begin{split}
&\mu= \left(\mu _0-\mu _1 r^{2 \alpha }\right)^{\beta}\,,\\
& p_r=\frac{2 b \delta  \mu _0}{ \mu _0
   \left(a+b r^2\right)}\\
&~~~~~~~- \frac{\left[a+b (2 \delta +1) r^2\right] \left(\mu _0-\mu _1
   r^{2 \alpha }\right)^{\beta +1}}{3 \mu _0
   \left(a+b r^2\right)}\mathcal{F}\,,\\
& p_\perp=\frac{2 \left[a^2+a b (2-3 \delta ) r^2+b^2 \left(-2
   \delta ^2+\delta +1\right) r^4\right]  \, \mathcal{F}}{\mu _012 \left(a+b r^2\right)^2\left(\mu _0-\mu _1 r^{2 \alpha
   }\right)^{-(\beta +1)}}\\
&+\frac{12 b \delta  \left(2 a+b \delta 
   r^2\right)-6 \left(a+b r^2\right) \left(a+b (\delta +1) r^2\right) }{12 \left(a+b r^2\right)^2\left(\mu _0-\mu _1
   r^{2 \alpha }\right)^{-\beta }}\,.
\end{split}
\end{equation}
Notice that the radial and tangential pressure converge to the same value in the centre of the star. The radial and tangential barotropic factor read
\begin{equation}\label{RecAni1wr}
\begin{split}
& w_r=\frac{\left(\mu _0-\mu _1 r^{2 \alpha }\right)}{3 \alpha  \beta  \mu _0 \mu _1 r^{2 \alpha } \left(a+b
   r^2\right)^2} \times\\
 &~~~~~~\left\{\frac{3}{2}
   \mu _0 \left(a+b r^2\right) \left[a+b (2 \delta +1) r^2\right]\right.\\
&~~~~~~~\-\frac{\mathcal{F}}{2} 
   \left(\mu _0-\mu _1 r^{2 \alpha }\right) \left[2 b \delta  r^2 \left(a+3 b
   r^2\right)+3 \left(a+b r^2\right)^2\right]\\
 &~~~~~+6 b^2 \delta  \mu _0 r^2 \left(\mu _0-\mu
   _1 r^{2 \alpha }\right){}^{-\beta }\Big\}\,,\\
  \end{split}
\end{equation}
 \begin{equation}\label{RecAni1worth}
\begin{split}
& w_\perp=\frac{r^{-2 \alpha }}{12 \alpha  \beta  \mu _1 \left(a+b r^2\right)^3} \times\\
&~~~~~~\left\{\frac{\mathcal{F}}{\mu _0} \left(\mu _0-\mu _1 r^{2 \alpha }\right){}^2
   \left[3 a^3-3 a^2 b (\delta -3) r^2 \right.\right.\\
&~~~~~~\left.\left.+a b^2 (2 (\delta -8) \delta +9) r^4+3 b^3 \left(-2
   \delta ^2+\delta +1\right) r^6\right]\right.\\
&~~~~~~~+12 b^2 \delta  r^2 \left[b \delta 
   r^2-a (\delta -4)\right] \left(\mu _0-\mu _1 r^{2 \alpha }\right){}^{1-\beta
   }\\
&~~~~~~~-3\mu _0 \left(a+b r^2\right) 
   \left[a^2+a b (2-5 \delta ) r^2\right.\\
&~~~~~~~ \left.+b^2 \left(-2 \delta ^2+\delta +1\right) r^4\right]\\
&~~~~~~~\left.-3\mu
   _1  r^{2 \alpha } \left(a+b r^2\right) \left[ab \delta  r^2  (2 \alpha  \beta +5)+2 b^2 \delta ^2
   r^4\right.\right.\\
&~~~~~~\left.\left.+b^2 \delta  r^4  (2 \alpha 
   \beta -1)+(2 \alpha  \beta -1) \left(a+b r^2\right)^2\right]\right\}\,.
\end{split}
\end{equation}
In Figs.~\ref{F:RecANI1a} and~\ref{F:RecANI1b}, we give a plot of the above solution for a set of values of the parameters which return physically relevant results. 

As a second example let us consider the case
\begin{equation}
\begin{split}\label{RecANIMetric2}
&A=\left(a+\left(c-b r^2\right)^{\gamma }\right)^{\delta}\,,\\
&B=\frac{3
   \mu _0}{3 \mu
   _0-r^2  \left(\mu _0-\mu _1 r^{2 \alpha }\right){}^{\beta +1}\mathcal{F}}\,,\\
&\mathcal{F}=\,
   _2F_1\left(1,\beta +\frac{3}{2 \alpha }+1;1+\frac{3}{2 \alpha };\frac{r^{2\alpha  \rho
   } \mu _1}{\mu _0}\right)\,,
\end{split}
\end{equation}
which can be seen as a generalisation of the  Bowers-Liang solution for non-constant densities.

\begin{figure}[!ht]
    \subfloat[A semilogarithmic plot of the coefficients of the metric \rf{RecANIMetric2}.The blue line represents $A$ and the orange $B$. \label{F:RecANIMetric2}]{%
      \includegraphics[width=0.45\textwidth]{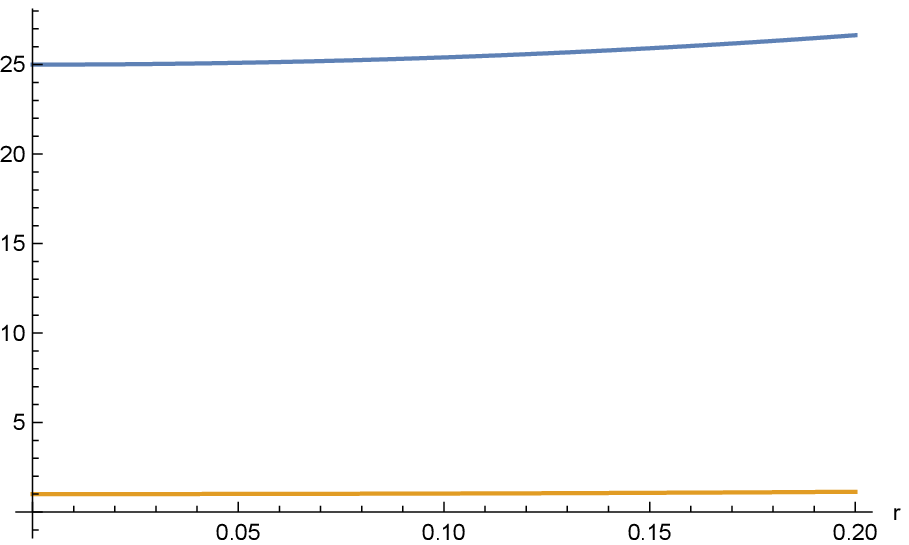}}\\
    \subfloat[A semilogarithmic plot of the thermodynamic quantities \rf{RecAni2Mu} and  \rf{RecAni2PP}  associated with \rf{RecANIMetric2}. The blue line represents  $p_r$, the orange $p_\perp$ and the green $\mu$.\label{F:RecAni2MuPP}]{%
      \includegraphics[width=0.45\textwidth]{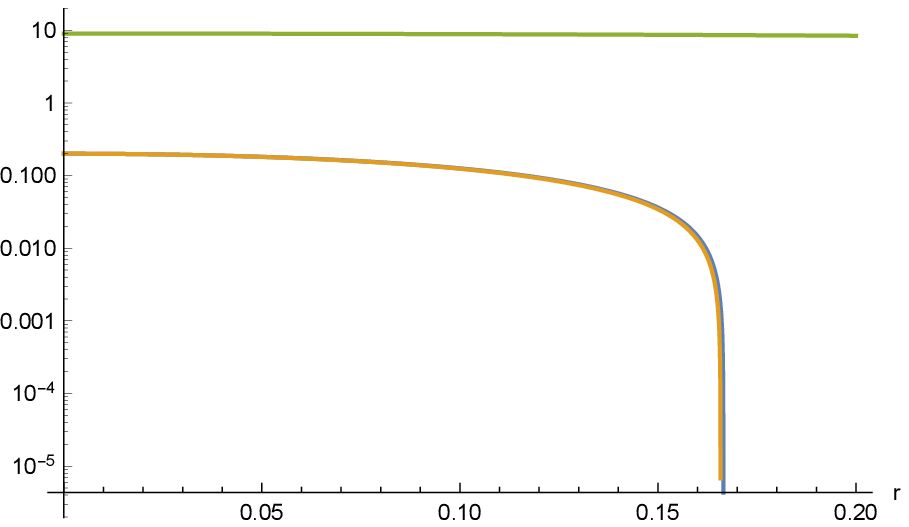}}\\
       \caption{Graphs of the solution \rf{RecANIMetric2} in the case $r_0=1$,  $\alpha=1$, $\beta=1$, $\gamma=2$, $\delta=2$, $a=10$, $b=-1$, $c=2$,  $\mu_0=9$, $\mu_1=15$, $A_0=1$. The values of the parameters have been chosen in order to make the features of the solution as clear as possible.}
          \label{F:RecANI2a}
  \end{figure}

\begin{figure}[!ht]
    \subfloat[The behaviour of the anisotropic pressure $\Pi$  associated with \rf{RecANIMetric2}.\label{F:RecAni2Pi}]{%
      \includegraphics[width=0.45\textwidth]{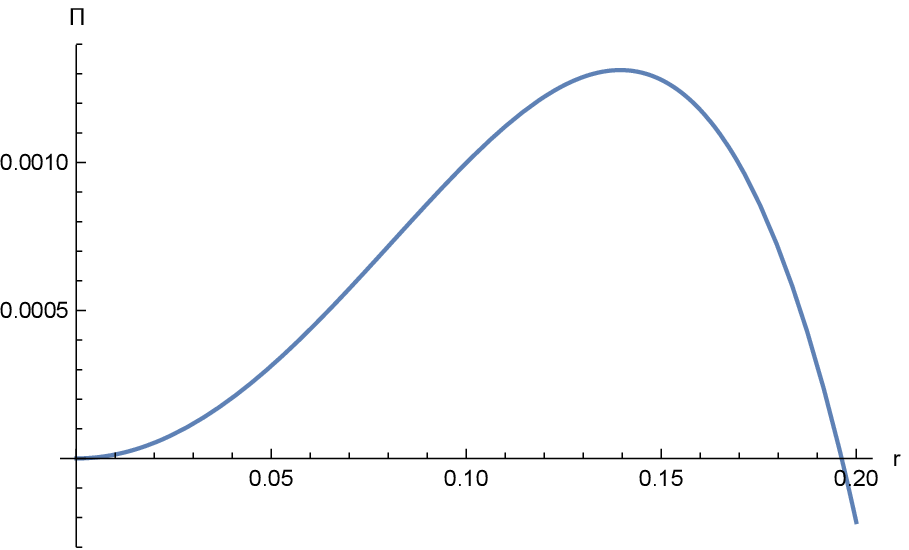}}\\
    \subfloat[The radial and orthogonal barotropic factors (blue and orange line, respectively) associated with \rf{RecANIMetric2}.\label{F:RecAni2ww}]{%
      \includegraphics[width=0.45\textwidth]{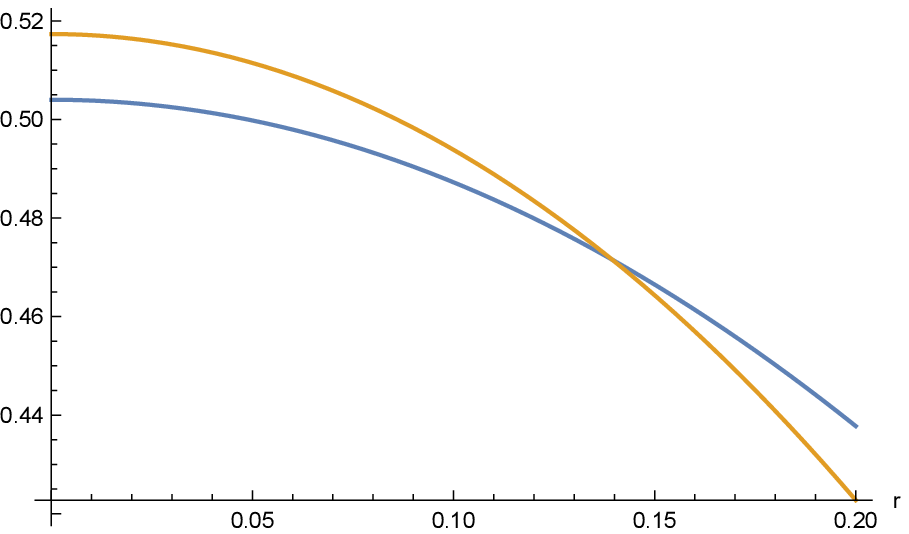}}\\
       \caption{Graphs of the solution \rf{RecANIMetric2} in the case $r_0=1$, $\alpha=1$, $\beta=1$, $\gamma=2$, $\delta=2$, $a=10$, $b=-1$, $c=2$,  $\mu_0=9$, $\mu_1=15$, $A_0=1$. The values of the parameters have been chosen in such a way to make the features of the solution as clear as possible.}
          \label{F:RecANI2b}
  \end{figure}

Setting for brevity  $r_0=1$, we have 
\be\label{RecAni2Mu}
\mu= \left(\mu _0-\mu _1 r^{2 \alpha}\right)^{\beta}\,,
\ee
and 
\begin{equation}\label{RecAni2P}
\begin{split}
& p_r=\frac{1}{6 \left(b r^2-c\right) \left(a+\left(c-b
   r^2\right)^{\gamma }\right)} \times\\
&~~~~~~\left\{12 b
   \gamma  \delta  \left(c-b r^2\right)^{\gamma }-2 \frac{\mathcal{F}}{\mu_0} \left[a \left(b r^2-c\right)\right.\right.\\
&~~~~~~\left.\left.+\left(c-b r^2\right)^{\gamma } \left(b r^2 (2 \gamma 
   \delta +1)-c\right)\right]\right\}\,,
  \end{split}
\end{equation}
\begin{equation}\label{RecAni2PP}
\begin{split}
& p_\perp=\frac{\mu _0^{\beta } \left(\mu _0-\mu
   _1 r^{2 \alpha }\right)^{-\beta }}{12 \left(b r^2-c\right)^2 \left(a+\left(c-b
   r^2\right)^{\gamma }\right)^2}\\
&~~~~~~\left\{2 \mathcal{F} \left(\mu _0-\mu _1 r^{2
   \alpha }\right)^{\beta } \left[a^2 \left(b r^2-c\right)^2
  \right.\right.\\
&~~~~~~\left.\left.
-a \left(c-br^2\right)^{\gamma } \left[b^2 r^4 \left(4 \gamma ^2 \delta -\gamma  \delta
   -2\right)  \right.\right.\right.\\
&~~~~~~\left.\left.\left.+b c r^2 (4-3 \gamma  \delta )-2 c^2\right]+\left(c-b r^2\right)^{2 \gamma }\times 
\right.\right.\\
&~~~~~~\left.\left.
   \left(b^2 r^4 \left(-2 \gamma ^2 \delta ^2+\gamma  \delta +1\right)+b c r^2 (3 \gamma 
   \delta -2)+c^2\right)\right]
\right.\\
&~~~~~~\left.
-3 \mu _0^{-\beta } \left(\mu _0-\mu _1 r^{2 \alpha
   }\right)^{\beta } \left[2 a^2 \left(b r^2-c\right)^2 \left(\mu _0-\mu _1 r^{2 \alpha
   }\right)^{\beta }
\right.\right.\\
&~~~~~~\left.\left.   
   +2 a \left(c-b r^2\right)^{\gamma } \left(b^2r^4 (\gamma
    \delta +2) \left(\mu _0-\mu _1 r^{2 \alpha }\right)^{\beta }
  \right.\right.\right.\\
&~~~~~~\left.\left.\left.
 -4 b^2\gamma ^2 \delta
  r^2  +b c \left(4 \gamma  \delta -r^2 (\gamma  \delta +4) \left(\mu _0-\mu _1 r^{2
   \alpha }\right){}^{\beta }\right)
   \right.\right.\right.\\
&~~~~~~\left.\left.\left.
   +2 c^2 \left(\mu _0-\mu _1 r^{2 \alpha}\right){}^{\beta }\right)+\left(c-b r^2\right)^{2 \gamma } \times
   \right.\right.\\
&~~~~~~\left.\left.
\left(b^2 r^2 \left(2 r^2
   (\gamma  \delta +1) \left(\mu _0-\mu _1 r^{2 \alpha }\right){}^{\beta }-4 \gamma ^2
   \delta ^2\right)
 \right.\right.\right.\\
&~~~~~~\left.\left.\left.   
   +2 b c \left(4 \gamma  \delta -r^2 (\gamma  \delta +2) \left(\mu
   _0-\mu _1 r^{2 \alpha }\right)^{\beta }\right)
 \right.\right.\right.\\
&~~~~~~\left.\left.\left.  
   +2 c^2 \left(\mu _0-\mu _1 r^{2 \alpha
   }\right){}^{\beta }\right)\right]\right\}\,.
  \end{split}
\end{equation}
Notice that the radial and tangential pressure converge to the same value in the centre of the star. The radial and orthogonal barotropic factors are too long to be reported here, but their calculation is trivial.
In Figs.~\ref{F:RecANI2a} and~\ref{F:RecANI2b}, we give a plot of the above solution for a set of values of the parameters fulfilling the physical criteria of Sec. \ref{ANISOTOV}. 
In light of the theorems that we have presented in the previous Section, this solution is somehow expected. It corresponds to the application of the theorem on the shift of the energy density, the isotropic pressure and the radial component of the metric.

We end this section pointing out an interesting aspect of the above solutions. At least in the cases that we have explored in Figs.~\ref{F:RecANI1a}   and~\ref{F:RecANI2a},  the radial and tangential pressure are relatively close to each other.  For an object of this type, therefore, the degree of anisotropy even if present is rather small. This fact points to the conclusion that a class of regular objects can exist which are {\it quasi-isotropic}.  Quasi isotropic stars  would behave like an isotropic star upon isolated observation, but they would differ dynamically because of the different properties of the anisotropic pressure. Such differences might appear evident and be studied, for example, at perturbative level.



\section{Conclusions}
In this paper we have used the $1+1+2$ covariant formalism and a tailored variable choice to develop a covariant version of the Tolman-Oppenheimer-Volkoff equations for objects that present the maximum degree of anisotropy compatible with spherical symmetry. Within this framework  it becomes clear that the anisotropy generates additional pressure terms which modify one of the isotropic TOV equations. This anisotropic pressure corresponds to the anisotropy term $\Delta$ which generally appears in literature.  

The covariant equations clarify the role of the anisotropic pressure in the structure of an object interior and immediately suggest a number of analytical resolution strategies. Indeed, some of these strategies have already been employed  in literature in specific cases like the Bowers-Liang one. 

The structure of the covariant TOV also suggests that there exists a number of algorithms which allow to map isotropic solutions in anisotropic ones. Indeed one of such methods has been recently proposed in Ref.~\cite{Ovalle:2017fgl}. These procedures can be useful to appreciate the physical role of the anisotropy in these systems as well as to explore further the solution space for anisotropic stars. We proposed some new algorithms of this type. In some cases these methods allow to recognise general properties of known solutions, like in the case of Florides' one.

In the isotropic case, a number of generating theorems were discovered which allow to connect different isotropic solutions to each other. The formalism that we have used allows to extend these theorems to the anisotropic case in a straightforward way. Indeed, we were able to find a number of new theorems that involve directly the anisotropic pressure.  As in the isotropic case we can therefore talk about seed metrics and organise the known solutions in terms of their relations via these theorems.

Finally, the new equations were used to derive a reconstruction algorithm able to generate anisotropic solutions. Unexpectedly this algorithm is much easier than its isotropic counterpart. In fact it allows to straightforwardly generate a double infinity of solutions. 

We used this algorithm to derive some new exact solutions. One of them is in a  generalisation of the Bowers-Liang solution for which the density is not constant. Its existence is expected by the generating theorems we have proven, but  a closer analysis revealed an unexpected feature: for some values of the parameters they show a radial and tangential pressure very close to each other.
Since the other solution we reconstructed  has the same properties one is lead to think that these ``quasi-isotropic relativistic stars''  might be a new class of objects never considered before.  It would be interesting to explore further and in more detail the properties of this class of objects. A future work will be focused specifically on this task.

\begin{acknowledgments}
SC and DV were supported by the Funda\c{c}\~{a}o para a Ci\^{e}ncia e Tecnologia through project IF/00250/2013 and acknowledge financial support provided under the European Union's H2020 ERC Consolidator Grant ``Matter and strong-field gravity: New frontiers in Einstein's theory'' grant agreement No. MaGRaTh646597.
\end{acknowledgments}

\end{document}